\documentclass[11pt]{article}
\usepackage{amssymb}

\newcommand{\bea}{\begin{eqnarray*}}
\newcommand{\eea}{\end{eqnarray*}}
\newcommand{\alfa}{\alpha}
\newcommand{\fie}{\varphi}

\newtheorem{teorema}{Theorem}
\newtheorem{lemma}{Lemma}
\newtheorem{opmerking}{Remark}

\begin{document}

\title{The Birkhoff theorem \\ for unitary matrices of prime-power dimension}
\author{Alexis De Vos and Stijn De Baerdemacker}

\maketitle

\begin{abstract}
The unitary Birkhoff theorem states that
any unitary matrix with all row sums and all column sums equal unity
can be decomposed as a weighted sum of permutation matrices, such that both 
the sum of                       the weights and 
the sum of the squared moduli of the weights are equal to unity.
If the dimension~$n$ of the unitary matrix equals 
a power of a prime~$p$, i.e.\  if $n=p^w$,
then the Birkhoff decomposition does
not need all $n!$~possible permutation matrices, as
the epicirculant permutation matrices suffice. 
This group of permutation matrices is
isomorphic to the general affine group GA($w,p$) of order
only $p^w(p^w-1)(p^w-p)...(p^w-p^{w-1}) \ll \left( p^w \right)!$.
\end{abstract}

\section{Introduction}

Let D($n$) be the semigroup of $n \times n$ doubly stochastic matrices;
let P($n$) be the     group of $n \times n$ permutation matrices.
Birkhoff \cite{birkhoff} has demonstrated 
\begin{teorema}
Every D($n$) matrix $D$ can be written 
       \[
       D = \sum_{\sigma} c_{\sigma} P_{\sigma}
       \]
       with all $P_{\sigma} \in$ P($n$) and the weights $c_{\sigma}$ 
       real, satisfying both $0 \le c_{\sigma} \le 1$
       and $\sum_{\sigma} c_{\sigma} = 1$. 
\end{teorema}
The question arises whether a similar theorem holds
for matrices from the unitary group U($n$).
This question is discussed by De Baerdemacker et al.\ 
\cite{debaerdemacker} \cite{devos}.
For this purpose, the subgroup XU($n$) of~U($n$) is introduced 
\cite{negator} \cite{subgroups}.
It consists of all U($n$) matrices with all line sums
(i.e.\ all row sums and all column sums) equal to~1.
Whereas U($n$) is an $n^2$-dimensional Lie group,
the group XU($n$) is only $(n-1)^2$-dimensional.
A unitary Birkhoff theorem has been proved 
for XU($n$) matrices \cite{debaerdemacker} \cite{devos}.
Remarkable is the fact that 
the case $n=p$ with $p$ an arbitrary prime \cite{devos}
has been treated in a very different way from 
the case where $n$ is an arbitrary integer \cite{debaerdemacker}.
As a result, the decomposition,
tailored to prime numbers \cite{devos}, can be restricted to $n^2$~terms,
whereas the general case \cite{debaerdemacker} leads to
a summation over all~$n!$ (or at least over~$n!/2$) permutation matrices,
albeit with a large number of degrees of freedom.
In the present paper, we will treat the two cases in a unified way.
Moreover, the unified approach will be applied to the
case $n=p^w$, i.e.\ $n$~equal to an arbitrary power~$w$ of an arbitrary prime~$p$.

In general, the Birkhoff theorem for unitary matrices
is easily proved as follows.
Let G($n$) be a finite subgroup of XU($n$).
\begin{lemma}
       If an XU($n$) matrix $X$ can be written 
       \[
       X = \sum_{\sigma} c_{\sigma} G_{\sigma}
       \]
       with all $G_{\sigma} \in$ G($n$), then the weights $c_{\sigma}$ 
       satisfy $\sum_{\sigma} c_{\sigma} = 1$. 
\end{lemma}
The proof is trivial: 
all line sums of $G_{\sigma}$ equal unity; therefore, 
all line sums of the matrix $c_{\sigma} G_{\sigma}$ equal $c_{\sigma}$ and thus 
all line sums of the matrix $\sum_{\sigma} c_{\sigma} G_{\sigma}$
are equal to $\sum_{\sigma} c_{\sigma}$.
As all line sums of $X$ are equal to~1,
we thus need $\sum_{\sigma} c_{\sigma} = 1$.

\begin{lemma}
       If every XU($n$) matrix $X$ can be written 
       \[
       X = \sum_{\sigma} a_{\sigma} G_{\sigma}
       \]
       with all $G_{\sigma} \in$ G($n$), then there exists a decomposition
       \[
       X = \sum_{\sigma} b_{\sigma} G_{\sigma} \ ,
       \]
       such that not only $\sum_{\sigma} b_{\sigma} = 1$, 
       but also $\sum_{\sigma} |b_{\sigma}|^2 = 1$.
\end{lemma}
This fact follows from the Klappenecker--R\"otteler theorem \cite{klap}.

\section{The group XU($n$)}

\begin{opmerking}
For sake of convenience,
in the present paper, the rows and colums of a matrix 
are not numbered starting from~1, but instead starting from~0.
Thus the upper-left entry of any $m \times m$ square matrix~$A$
is $A_{0,0}$ and its lower-right entry is $A_{m-1,m-1}$.
\end{opmerking}

We recall that
the group XU($n$) is an $(n-1)^2$-dimensional subgroup of the $n^2$-dimensional
unitary group~U($n$).
Any member $X$ of XU($n$) can be written
\begin{equation}
X = T\ \left(\begin{array}{cc} 1 & \\ & U \end{array} \right)\ T^{-1} \ ,
\label{TUT}
\end{equation}
where $U$ is a member of U($n-1$) and
where the constant unitary matrix~$T$ is $1/\sqrt{n}$ times
a dephased complex Hadamard matrix \cite{tadej}.
Thus (\ref{TUT}) constitutes a 1-to-1 mapping between~$X$ and~$U$.
Because of 
\begin{equation}
T_{j,0}=T_{0,k}=1/\sqrt{n} \ ,
\label{11}
\end{equation}
eqn (\ref{TUT}) leads to
\[
X_{k,l} = \frac{1}{n}\ + \sum_{r=1}^{n-1} \sum_{s=1}^{n-1} 
         T_{k,r}\, U_{r-1,s-1}\, (T^{-1})_{s,l} \ .
\]
With $T$ being unitary, i.e.\ with $T^{-1}=T^{\dagger}$, this becomes
\[
X_{k,l} = \frac{1}{n}\ + \sum_{r=1}^{n-1} \sum_{s=1}^{n-1} 
         \, U_{r-1,s-1}\, T_{k,r}\overline{T_{l,s}} \ .
\]
We thus can write the matrix~$X$ as a sum of $1 + (n-1)^2$ matrices:
\begin{equation}
X = W + \frac{1}{n}\ \sum_{r=1}^{n-1} \sum_{s=1}^{n-1} U_{r-1,s-1}M_{r,s} \ ,
\label{XWUM}
\end{equation}
where $W$ is the $n \times n$ van der Waerden matrix,
i.e.\ the doubly stochastic matrix with all entries equal to $\frac{1}{n}$, and
where $M_{r,s}$ is an $n \times n$ matrix defined by
\begin{equation}
(M_{r,s})_{k,l} = n\, T_{k,r}\overline{T_{l,s}} \ .
\label{Mrskl}
\end{equation}
The  labels~$r$ and~$s$ of the matrix $M_{r,s}$ run from~1 to~$n-1$, in contrast to 
the indices~$k$ and~$l$ of its entries, which   run from~0 to~$n-1$.
We thus have $(n-1)^2$ such matrices, each having $n^2$~entries.
Each entry of the matrix $M_{r,s}$ equals 
the  leftmost entry of its row times
the uppermost entry of its column.
Taking into account (\ref{11}),
one indeed easily checks
\begin{equation}
(M_{r,s})_{0,l}\ (M_{r,s})_{k,0} = (M_{r,s})_{k,l} \ .
\label{MMM}
\end{equation}
Both the first row and the first column of $M_{r,s}$ 
equal a line of the Hadamard matrix $T$ 
(up to complex conjugation and up to the factor $\sqrt{n}\,$):
\begin{eqnarray}
(M_{r,s})_{0,l} & = & \sqrt{n}\ \overline{T_{l,s}}    \nonumber \\
(M_{r,s})_{k,0} & = & \sqrt{n}\           T_{k,r} \ . \label{1}
\end{eqnarray} 
Because $T$ is $1/\sqrt{n}$ times a Hadamard matrix, 
we have $|T_{l,s}|=1/\sqrt{n}$ and $|T_{k,r}|=1/\sqrt{n}$,
such that $|(M_{r,s})_{0,l}|=1$ and $|(M_{r,s})_{k,0}|=1$, and thus, because of (\ref{MMM}),
we conclude that all entries $(M_{r,s})_{k,l}$ have unit modulus.

\section{Underlying framework}

In the present section, we consider an arbitrary doubly transitive group~G($n$)
of $n \times n$ permutation matrices. 
We denote by $N$ the order of the group.
We generalize the ideas and computations
in Reference \cite{debaerdemacker}, where G($n$) is 
equal to the group P($n$) of all $n \times n$ permutation matrices,
thus G($n$) being isomorphic to the symmetric group {\bf S}$_n$
and $N$ being equal to~$n!$.

In the next three sections, we will apply the Lemmas~1 and~2 to
three different choices of G($n$):
\begin{itemize}
\item In case of arbitrary~$n$, 
      we choose the group of all $n \times n$ permutation matrices
      (i.e.\ a group isomorphic to the symmetric group {\bf S}$_n$).
      See Section~4.
\item In case of $n$ equal to some prime~$p$, 
      we choose the group of all $n \times n$ supercirculant permutation matrices
      (i.e.\ a group isomorphic to a semidirect-product group {\bf C}$_n$~{\bf : C}$_{n-1}$).
      See Section~5.
\item In case of $n$ equal to some power $w$ of some prime~$p$
      (i.e.\ equal to $p^w$), 
      we choose the group of all $n \times n$ epicirculant permutation matrices
      (i.e.~a group isomorphic to the general affine group GA($w,p$)).
      See Section~6.
\end{itemize}
The meaning of the words `supercirculant' and `epicirculant' will be made clear below.
The mentioned groups are doubly transitive, as it is known that
the symmetric   group {\bf S}$_n$ is     $n$-transitive, 
the alternating group {\bf A}$_n$ is $(n-2)$-transitive, and 
the affine groups are $2$-transitive \cite{wolfram}, in contrast to e.g.\ 
the cyclic      group {\bf C}$_n$, which is only 1-transitive.
 
In each of the three cases, we will prove below 
that every XU($n$) matrix $X$ can be written as 
       \begin{equation}
       X = \sum_{\sigma} c_{\sigma} G_{\sigma}
       \label{G}
       \end{equation}
       with all $G_{\sigma}$ member of the appropriate group~G($n$).
Because of Lemmas~1 and~2, we are then allowed to put the case
that both $\sum_{\sigma} c_{\sigma} = 1$ 
and $\sum_{\sigma} |c_{\sigma}|^2 = 1$.
For the explicit computation of the weights~$c_{\sigma}$,
we note that the G($n$) matrices form
an $n$-dimensional reducible representation of some abstract group~{\bf G}.
We assume that {\bf G} has $\mu$ different irreducible representations.
According to Lemma~(29.1) of \cite{burrow},
because {\bf G} is 2-transitive, the $n$-dimensional natural representation
is the sum of the 1-dimensional trivial representation
and an $(n-1)$-dimensional irreducible representation, 
which we will call the standard representation.

We replace eqn (\ref{G}) by an eqn 
concerning one of the $\mu$~irreducible representations of~{\bf G}: 
       \begin{equation}
       U^{(\nu)} = \sum_{\sigma} c_{\sigma} D^{(\nu)}_{\sigma} \ ,
       \label{D}
       \end{equation}
where $\nu$ is the label of the irrep ($0 \le \nu \le \mu -1$),
where $D^{(\nu)}_{\sigma}$ is the $\nu$~th
irreducible representation of $G_{\sigma}$, and $U^{(\nu)}$ is an
appropriate $n_{\nu} \times n_{\nu}$ unitary matrix,
with a special mentioning for $\nu=0$ anf $\nu=1$ (see further).
Here, $n_{\nu}$ is the dimension of the $\nu$~th representation.
We have $\mu$ such matrix equations~(\ref{D}). 
Each matrix eqn constitutes $n_{\nu}^2$ scalar equations. 
We thus have a total of
$\sum_{\nu=0}^{\mu-1} n_{\nu}^2 = N$ scalar equations with $N$ unknowns~$c_{\sigma}$:
\[
\sum_{\sigma} c_{\sigma} (D^{(\nu)}(\sigma))_{k,l} = (U^{(\nu)})_{k,l} \ . 
\]
Solution of this set of equations is:
\begin{eqnarray}
c_{\sigma} & = & \frac{1}{N} \sum_{\nu} n_{\nu} 
                 \sum_{i=0}^{n_{\nu}-1} \sum_{j=0}^{n_{\nu}-1} 
                 (D^{(\nu)}(\sigma))_{i,j}\ (U^{(\nu)}(\sigma))_{i,j} \nonumber \\ 
           & = & \frac{1}{N} \sum_{\nu} n_{\nu}\ \mbox{Tr}
                 \left( D^{(\nu)}(\sigma)^{\dagger}\, U^{(\nu)}(\sigma)\right) \ . 
\label{formule}
\end{eqnarray}

We choose for $\nu=0$ the trivial representation,
i.e.\ the 1-dimensional irreducible representation with all characters equal to~1.
We choose for $\nu=1$ the standard representation,
i.e.\ the $(n-1)$-dimensional irreducible representation obtained by 
applying (\ref{TUT}) to the permutation matrix $P_{\sigma}$:
\[
P_{\sigma} = T\ \left(\begin{array}{cc} 1 & \\ & D^{(1)}(\sigma) \end{array} \right)\ T^{-1}
\]
and thus 
\[
\left(\begin{array}{cc} 1 & \\ & D^{(1)}(\sigma) \end{array} \right) = T^{-1} P_{\sigma} T \ .
\]

In (\ref{formule}), 
the matrix $U^{(0)}(\sigma)$ equals the $1 \times 1$ unit matrix and 
the matrix $U^{(1)}(\sigma)$ equals the $(n-1) \times (n-1)$ lower-right block of
\[
\left(\begin{array}{cc} 1 & \\ & U \end{array} \right) = T^{-1} X T \ .
\]
For the remaining matrices $U^{(\nu)}(\sigma)$ with $2 \le \nu \le \mu -1$,
we are allowed to choose any unitary matrix of the right dimension~$n_{\nu}$.
This usually allows a large number of degrees of freedom.
Here, we propose two different strategies to take advantage of this freedom.

\subsection{First strategy}

For each matrix $U^{(\nu)}(\sigma)$ with $2 \le \nu \le \mu -1$,
we choose the $n_{\nu} \times n_{\nu}$ unit matrix.
Then (\ref{formule}) becomes
\begin{equation}
c_{\sigma} = \frac{1}{N}\ 
             [\                  n_{0}  \, \mbox{Tr}\left(D^{(0)  \, \dagger}(\sigma)    \right) + 
                                 n_{1}  \, \mbox{Tr}\left(D^{(1)  \, \dagger}(\sigma)U   \right) + 
              \sum_{\nu=2}^{\mu -1} n_{\nu}\, \mbox{Tr}\left(D^{(\nu)\, \dagger}(\sigma) \right) \ ] \ .
\label{formule1}
\end{equation} 
We take advantage of Shur's orthogonality relation:
\[
 \sum_{\nu} n_{\nu}\,  \mbox{Tr}\left(D^{(\nu)\, \dagger}(\sigma)\right) = 
 \sum_{\nu} n_{\nu}\,  \mbox{Tr}\left(D^{(\nu)\, \dagger}(\sigma)\, D^{(\nu)}(\epsilon)\right) =
 \delta_{\sigma}\, N \ ,
\]
where $\epsilon$ is the trivial identity permutation and
where $\delta_{\epsilon}=1$ while $\delta_{\sigma}=0$ if $\sigma \neq \epsilon$.
Because moreover $D^{(1)\, \dagger}(\sigma) = D^{(1)}(\sigma^{-1})$
and $n_1=n-1$, we obtain the explicit expression for the weight:
\begin{equation}
c_{\sigma} = \delta_{\sigma}+ 
             \frac{n-1}{N}\ \mbox{Tr}\left(D^{(1)}(\sigma^{-1})U \right)  - 
             \frac{n-1}{N}\ \chi^{(1)}(\sigma^{-1})  \ .  
\label{chi} 
\end{equation} 
The number $\chi^{(\nu)}(G)$ denotes the character of the element $G$ of the group~{\bf G}
according to the $\nu$~th representation. 
It is equal to $\mbox{Tr}(D^{(\nu)}(G))$. 
In particular, we have $\mbox{Tr}(D^{(1)}(G))= \mbox{Tr}(G)-1$.  

\subsection{Second strategy}

The second strategy is only applicable if the group~{\bf G} has
an anti-standard irreducible representation,
non-equivalent to the standard representation.
The anti-standard representation, 
which we will assign the label~$\nu=2$ (if it exists),
has the same characters as the standard representation (with label~$\nu=1$),
except for a factor~$-1$ if the corresponding permutation 
is an odd permutation.
A necessary condition for the second strategy is
\begin{equation}
N \ge 2 + 2(n-1)^2 \ .
\label{N>}
\end{equation} 

As in the first strategy, we again choose the $1 \times 1$ unit matrix for $U^{(0)}(\sigma)$
and the $(n-1) \times (n-1)$ matrix $U$ for $U^{(1)}(\sigma)$.
However, in this second strategy, we also choose the matrix~$U$ for
each matrix $U^{(2)}(\sigma)$. 
For each matrix $U^{(\nu)}(\sigma)$ with $3 \le \nu \le \mu -1$,
we choose the $n_{\nu} \times n_{\nu}$ unit matrix.
Then (\ref{formule}) becomes
\begin{eqnarray}
c_{\sigma} = \frac{1}{N}\ 
             [\  n_{0}  \, \mbox{Tr}\left(D^{(0)  \, \dagger}(\sigma) \right)              & + & 
                 n_{1}  \, \mbox{Tr}\left(D^{(1)  \, \dagger}(\sigma) \right)                +
                 n_{2}  \, \mbox{Tr}\left(D^{(2)  \, \dagger}(\sigma) \right) \nonumber \\ & + &
                   \sum_{\nu=3}^{\mu -1} 
                 n_{\nu}\, \mbox{Tr}\left(D^{(\nu)\, \dagger}(\sigma) \right) \ ] \ .
\label{formule2}
\end{eqnarray} 
Again taking advantage of Shur's orthogonality relation and $n_1=n_2=n-1$, we obtain
\begin{eqnarray}
c_{\sigma} = \delta_{\sigma} & + \, 
             \frac{2(n-1)}{N}\ \mbox{Tr}\left(D^{(1)}(\sigma^{-1})U \right)  - 
             \frac{2(n-1)}{N}\ \chi^{(1)}(\sigma^{-1})  & \mbox{ if $\sigma$ even} \nonumber \\
           = \  0 & \ & \mbox{ if $\sigma$ odd}\ .  
\label{chi0}
\end{eqnarray}

In the second strategy, 
the group {\bf G}~$\cap$~{\bf A}$_n$ thus takes over the role of {\bf G} and 
$N/2$ takes over the role of~$N$.

\section{The case of arbitrary dimension $n$}

\begin{lemma}
Every XU($n$) matrix $X$ can be written 
       \[
       X = \sum_{\sigma} c_{\sigma} P_{\sigma}
       \]
       with all $P_{\sigma} \in$ P($n$).
\end{lemma}
The proof is provided by \cite{devos},
by means of induction on~$n$. 
Combining Lemmas~1, 2, and~3 leads to the unitary Birkhoff theorem:
\begin{teorema}
Every XU($n$) matrix $X$ can be written 
       \[
       X = \sum_{\sigma} c_{\sigma} P_{\sigma}
       \]
       with all $P_{\sigma} \in$ P($n$), such that both
       $\sum_{\sigma} c_{\sigma} = 1$ 
       and $\sum_{\sigma} |c_{\sigma}|^2 = 1$.
\end{teorema}

\subsection{First strategy}

We can apply result (\ref{chi}) with $N=n!$.
The only possible values of $\chi^{(1)}$ are Tr(P$_{\sigma}$)$-1$
and thus $-1, 0, 1, 2, ..., n-1$,
with exception of $n-2$. 

\subsection{Second strategy}

The character tables of the groups~{\bf S}$_2$ and~{\bf S}$_3$
show no anti-standard representation. 
For $n>3$, the group {\bf S}$_n$ has an anti-standard representation. 
In this case, we can apply result (\ref{chi0}) with $N=n!$.
The restriction $n>3$ is not surprising, as (\ref{N>}) with $N=n!$
is fulfilled neither if $n=2$ nor if $n=3$.

\section{The case of prime dimension $n=p$}

We call an $n \times n$ matrix $A$ supercirculant iff each row~$k$ equals row~$k-1$
shifted $x$~positions to the right. 
Thus $A_{k,l}=A_{k-1,l-x}$, where addition and subtraction are modulo~$n$.
We equivalently may write
\[
A_{0,a}=A_{k,a+kx} \ .
\]
We call $x$ the pitch of the matrix. 
If $x=1$,   then the supercirculant matrix is called     circulant;
if $x=n-1$, then the supercirculant matrix is called anticirculant.

If $p$ denotes a prime, then
the $p \times p$ supercirculant permutation matrices are denoted $S_{a,x}$,
where $x$ is the pitch and
$a$ (called the shift) is the column with the unit entry in the upper row (i.e.\ row~0).
The unit entries of such $p \times p$ permutation matrix
thus are located at the $p$~positions
$(0,a)$, $(1,a+x)$, $(2,a+2x)$, ..., and $(p-1,a+(p-1)x)$,
where sums are to be taken modulo~$p$. 
Because~$x$ and~$p$ are co-prime, the consecutive columns with a~1, i.e.\  
the columns $a$, $a+x$, $a+2x$, ..., and $a+(p-1)x$, are all different.

If $n$ equals some prime $p$, then
we choose for the $p \times p$ Hadamard matrix~$T$ of Section~2
the $p \times p$ discrete Fourier transform~$F$, with entries
\[
F_{k,l} = \frac{1}{\sqrt{p}}\ \omega^{kl} \ ,
\]
where $\omega$ is equal to the $p\,$th root of unity. Thus (\ref{Mrskl}) becomes
\[
(M_{r,s})_{k,l} = \omega^{kr-ls} \ .
\]
From \cite{devos}, we know that
$M$ can be written as a weighted sum of $p$ supercirculant permutation matrices:
\begin{equation}
M_{r,s} = \sum_{a=0}^{p-1}\, (M_{r,s})_{0,a}\ S_{a,x(r,s)} \ ,
\label{sum1}
\end{equation}
where the pitch~$x$ of the matrix $S_{a,x}$ is a function of $r$ and $s$.
Indeed, the condition
\[
(M_{r,s})_{k,a+kx} = (M_{r,s})_{0,a} 
\] 
yields
\[
kr - (a+kx)s = - as 
\]
and thus $r - xs = 0$.
Thus $x$~has to satisfy the eqn
\[
sx = r \ \mbox{mod}\ p \ .
\] 
This eqn has one solution: 
\[
x = rs^{-1} \ \mbox{mod}\ p \ ,
\] 
where $s^{-1}$ is the inverse of~$s$ modulo~$p$.
As~$p$ is prime, each non-zero integer has exactly one inverse.
With $(M_{r,s})_{0,a} = \omega^{-as}$, we finally obtain
\[
M_{r,s} = \sum_{a=0}^{p-1}\, \omega^{-as} S_{a,rs^{-1}} \ .
\]

The supercirculant 
$p \times p$ permutation matrices form a group S($p$),
subgroup of~P($p$) (proof in Appendix~A),
isomorphic to the semidirect product of the cyclic group of order~$p$ and
the multiplicative group of integers modulo~$p$.
The group thus is isomorphic to the semidirect product of two cyclic groups:
\[
{\bf C}_p \, \mbox{\bf :}\,  {\bf C}_{p-1} \ , 
\]
a non-Abelian group of order $p(p-1)$.
\begin{lemma}
If $n$ is prime, then
every XU($n$) matrix $X$ can be written 
       \[
       X = \sum_{\sigma} c_{\sigma} S_{\sigma}
       \]
       with all $S_{\sigma} \in$ S($n$).
\end{lemma}
The proof is as follows. 
If $n$ is a prime $p$, then all matrices $M_{r,s}$
are supercirculant with a pitch~$x=rs^{-1}$ modulo~$p$.
Also the van der Waerden matrix $W$ is supercirculant, as it is circulant:
\[
W = \sum_{a=0}^{n-1} \frac{1}{n}\ S_{a,1} \ .
\]
Hence, according to (\ref{XWUM}), $X$ is a weighted sum of 
supercirculant permutation matrices.
 
Combining Lemmas~1, 2, and~4 leads to
\begin{teorema}
If $n$ is prime, then
every XU($n$) matrix $X$ can be written 
       \[
       X = \sum_{\sigma} c_{\sigma} S_{\sigma}
       \]
       with all $S_{\sigma} \in$ S($n$), such that both
       $\sum_{\sigma} c_{\sigma} = 1$ 
       and $\sum_{\sigma} |c_{\sigma}|^2 = 1$.
\end{teorema}

\subsection{First strategy}
 
We can apply result (\ref{chi}) with $N=p(p-1)$.
The only possible values of $\chi^{(1)}$ are $-1$, $0$, and $p-1$,
as demonstrated in Appendix~B.
Thus we find a unitary Birkhoff decomposition with only $p(p-1)$ terms. 
For a prime exceeding~3, 
this number is substantially smaller than the number~$p!/2$ of Subsection~4.2.
The resulting unitary Birkhoff theorem is also slightly
stronger than the theorem in \cite{devos},
where the Birkhoff decomposition consists of $p^2$ terms.

\subsection{Second strategy}

The group S(2), isomorphic to the cyclic group {\bf C}$_2$,
has only two irreducible representations: the trivial one and the standard one.
Also the group S($n$) with $n$ equal to an odd prime~$p$,
has no inequivalent anti-standard representation.
Indeed, because all odd supercirculant permutations 
have non-unit pitch (see Appendix~C) and thus 
have unit trace (see Appendix~B) and hence 
have zero character $\chi^{(1)}$, all characters of the
anti-standard representation equal the corresponding characters
of the standard representation. Therefore,
the standard and anti-standard representations are equivalent.
We conclude that we cannot apply the second strategy of Subsection~3.2.
The absence of any inequivalent anti-standard representation is no surprise, 
as $N=n(n-1)$ does not satisfy (\ref{N>}).

\section{The case of prime-power dimension $n=p^w$}

For $n=p^w$ with arbitrary positive~$w$, 
we can choose for $T$ of Section~2 the Kronecker product
of $w$ small (i.e. $p \times p$) Fourier matrices~$F$:
\[
T = F \otimes F \otimes ... \otimes F = \ F^{\otimes w} \ .
\]
The $n \times n$ matrix~$T$ has following entries:
\[
T_{a,b} = \frac{1}{\sqrt{n}}\ \omega^{f(a,b)} \ ,
\]
where $f(x,y)$ is the sum of the ditwise product of the $p$-ary numbers~$x$ and~$y$:
\[
f(x,y) = \sum_j x_jy_j \mbox{ mod } p \ .
\]
As a consequence, we have
\begin{equation}
(M_{r,s})_{k,l} = \omega^{f(k,r) - f(s,l)} \ .
\label{epis}
\end{equation}
Among the $n^2$~entries of this matrix,
$n^2/p$ are equal to~1,
$n^2/p$ are equal to~$\omega$, ..., and
$n^2/p$ are equal to~$\omega^{p-1}$.

\begin{opmerking}
For sake of convenience, below, the rows and the colums of a matrix
will sometimes be pointed at, not by a number, but instead by a vector.
This will allow matrix computations for the row and column numbers.
For this purpose, any number
$z = z_0 + z_1p + z_2p^2 ... + z_{w-1}p^{w-1}$ has an
associated boldfaced $w \times 1$ vector
${\bf z} = (z_0, z_1, z_2, ..., z_{w-1})^T$,
consisting of the $w$~dits of the number~$z$. 
\end{opmerking}

We call a matrix $A$ epicirculant if row~$k$ equals row~0,
`shifted to the right' according to
\[
A_{{\bf 0},{\bf a}} = A_{{\bf k},\, {\bf a} + {\bf xk}} \ ,
\]
where {\bf a} is the $w \times 1$ vector associated with the column number~$a$ and
where {\bf x} is a   $w \times w$ matrix called the pitch matrix,
      consisting of $w^2$ entries, all $\in \{ 0, 1, ..., p-1\}$.
A matrix of the form (\ref{epis}) is automatically epicirculant.
It is a weighted sum of epicirculant permutation matrices $E$:
we have 
\begin{equation}
M_{r,s} = \sum_{a=0}^{p-1}\, (M_{r,s})_{0,a}\ E_{{\bf a},{\bf x}(r,s)} \ .
\label{ME}
\end{equation}
Here, ${\bf x}$ is an appropriate $w \times w$ pitch matrix, depending on~$r$ and~$s$. 
Proof is in Appendix~D.
We note that vector~${\bf a}$ and matrix~${\bf x}$ constitute a pair,
fully specifying an affine transformation \cite{wiki}.
  
If $n$ is a prime power, say $n=p^w$, then the epicirculant 
$p^w \times p^w$ permutation matrices form a group E($n$),
subgroup of~P($n$) (proof in Appendix~E),
isomorphic to the general affine group GA($w,p$),
a semidirect product of the direct product
of cyclic groups of order~$p$ and
the general linear group GL($w,p$):
\[
\mbox{GA}(w, p) = {\bf C}_p^w\, \mbox{\bf :} \, \mbox{GL}(w, p)
\] 
of order
\begin{equation}
p^w(p^w-1)(p^w-p)(p^w-p^2)...(p^w-p^{w-1}) \ .
\label{orde}
\end{equation}
We note that GA($w,p$) is a maximal subgroup of the symmetric group
{\bf S}$_{p^w}$ (O'Nan--Scott theorem) \cite{liebeck}.

Each of the $w$~subgroups {\bf C}$_p$ consists of $p$~matrices,
each a Kronecker product with a total of $w$~factors:
\[
I \otimes I \otimes ... \otimes I \otimes M \otimes I \otimes ... \otimes I \ ,
\]
where $I$ denotes the $p \times p$ unit                  matrix 
and   $M$ a           $p \times p$ circulant permutation matrix~$S_{a,1}$. 

\begin{lemma}
If $n$ is a prime power, then
every XU($n$) matrix $X$ can be written 
       \[
       X = \sum_{\sigma} c_{\sigma} E_{\sigma}
       \]
       with all $E_{\sigma} \in$ E($n$).
\end{lemma}
The proof is as follows.
If $n$ is a prime power $p^w$, then all matrices $M_{r,s}$
are epicirculant with an invertible pitch matrix {\bf x}.
Also the van der Waerden matrix $W$ is epicirculant, as it is circulant:
\[
W = \sum_{a=0}^{n-1} \frac{1}{n}\ E_{{\bf a},{\bf 1}} \ ,
\]
where the pitch matrix ${\bf 1}$ denotes the $w \times w$ unit matrix.
Hence, according to (\ref{XWUM}), $X$ is a weighted sum of 
epicirculant permutation matrices.
 
Combining Lemmas~1, 2, and~5 leads to
\begin{teorema}
If $n$ is a prime power, then
every XU($n$) matrix $X$ can be written 
       \[
       X = \sum_{\sigma} c_{\sigma} E_{\sigma}
       \]
       with all $E_{\sigma} \in$ E($n$), such that both
       $\sum_{\sigma} c_{\sigma} = 1$ 
       and $\sum_{\sigma} |c_{\sigma}|^2 = 1$.
\end{teorema}

\subsection{First strategy}

We can apply result (\ref{chi}) with $N$ given by (\ref{orde}).
The only possible values of $\chi^{(1)}$ are $-1$, $0$, $p-1$,
$p^2-1$, $p^3-1$, ..., and $p^w-1$,
as demonstrated in Appendix~F. 

\subsection{Second strategy}

For $w>1$ and $p>2$, the general affine groups have, 
besides the standard representation, also an inequivalent anti-standard representation.
For a proof, it suffices to point to a single example
of an odd epicirculant permutation matrix with trace different from unity.
We choose the $p^w \times p^w$ matrix
\[
E = I \otimes I \otimes ... \otimes I \otimes M \ ,
\]
i.e.\ the Kronecker product of $w-1$ matrices $I$ (i.e.\ the $p \times p$ unit matrix)
and the $p \times p$ supercirculant matrix $M = S_{\, 0,q}$.
The $w \times w$ pitch matrix associated with $E$
is the diagonal matrix $\mbox{diag}(q, 1, 1, ...,1)$.

On the one hand,
we have the following property of the Kronecker product
of two square matrices:
\begin{equation}
\mbox{Det}(A \otimes B) = [\, \mbox{Det}(A)\, ]^{\mbox{dim}(B)} \ 
                          [\, \mbox{Det}(B)\, ]^{\mbox{dim}(A)} \ .
\label{ABAB}
\end{equation}
Therefore, we have $\mbox{Det}(E) = \mbox{Det}(M)^{(p^{w-1})}$.
We choose the number~$q$ such that
$\mbox{Det}(M)=-1$ and thus $\mbox{Det}(E)=-1$.
This is always possible. Suffice it to choose $q$ equal to $g(p)$,
where $g$ is a generator of the modulo~$p$ multiplication group \cite{wolfram2}.
Unfortunately, there is no algorithm known for finding
such generator except brute force \cite{conrad}.
Nevertheless, we can prove that Det$(S_{0,g(p)}) = -1$, 
without a~priori knowing the value of $g(p)$: see Appendix~C.

On the other hand, 
we have $\mbox{Tr}(E) = p^{w-1}\, \mbox{Tr}(M) = p^{w-1}\, 1 = p^{w-1}$.
Because $w>1$, we have $\mbox{Tr}(E)>1$ and thus $\chi^{(1)}>0$. 
We thus conclude that we can apply result (\ref{chi0}) with $N$ according to (\ref{orde}).

The above reasoning is not valid for $p=2$, because, in that case,
$\mbox{Det}(M)=-1$ does not imply $\mbox{Det}(E)=-1$.
For the case $p=2$, we will prove that all $2^w \times 2^w$ epicirculant matrices
are even permutations. 
For this purpose, it is sufficient to demonstrate 
that all group generators are even.
From reversible computing \cite{beth} \cite{patel} \cite{boek}, it is known that the group
GA($w,2$) is generated by following matrices:
\bea
A & = & I \otimes I \otimes ... \otimes I \otimes 
        \left( \begin{array}{cc} 0 & 1 \\ 1 & 0 \end{array} \right) 
        \otimes I \otimes I \otimes ... \otimes I \\[1mm]
B & = & I \otimes I \otimes ... \otimes I \otimes 
        \left( \begin{array}{cccc} 1 & 0 & 0 & 0 \\ 
                                  0 & 1 & 0 & 0 \\
                                  0 & 0 & 0 & 1 \\
                                  0 & 0 & 1 & 0 \end{array} \right) 
        \otimes I \otimes I \otimes ... \otimes I \\[1mm]
C & = & I \otimes I \otimes ... \otimes I \otimes 
        \left( \begin{array}{cccc} 1 & 0 & 0 & 0 \\ 
                                  0 & 0 & 0 & 1 \\
                                  0 & 0 & 1 & 0 \\
                                  0 & 1 & 0 & 0 \end{array} \right) 
        \otimes I \otimes I \otimes ... \otimes I \ , 
\eea
with a total of $w-1$ (for $A$) or $w-2$ (for $B$ and $C$) factors~$I$.
In the context of computing, these matrices
represent {\tt NOT} gates, respectively controlled {\tt NOT} gates.
Applying (\ref{ABAB}), we have:
\bea
\mbox{Det}(A) & = & [\ \mbox{Det}\left( \begin{array}{cc} 0 & 1 \\ 1 & 0 \end{array} \right)\, ]^{(p^{w-1})} 
                = (-1)^{2^{w-1}} = 1 \\[1mm]
\mbox{Det}(B) & = & [\ \mbox{Det}\left( \begin{array}{cccc} 1 & 0 & 0 & 0 \\ 
                                                            0 & 1 & 0 & 0 \\
                                                            0 & 0 & 0 & 1 \\
                                                            0 & 0 & 1 & 0 \end{array} \right)\, ]^{(p^{w-2})}
                = (-1)^{2^{w-2}} = 1 \\[1mm]
\mbox{Det}(C) & = & [\ \mbox{Det}\left( \begin{array}{cccc} 1 & 0 & 0 & 0 \\ 
                                                            0 & 0 & 0 & 1 \\
                                                            0 & 0 & 1 & 0 \\
                                                            0 & 1 & 0 & 0 \end{array} \right)\, ]^{(p^{w-2})}
                = (-1)^{2^{w-2}} = 1 \ ,
\eea
except if $w=2$.
Thus, for $w>2$, all members of GA($w,2$) represent even permutations and 
the second strategy (Subsection~3.2) is not applicable.

This leaves us with the case $p=2$ and $w=2$. 
The epicirculant matrices form a group E(4) isomorphic to
the symmetric group~{\bf S}$_4$. As stated in Section~4.2, 
the second strategy is applicable. The results on the
applicability of the second strategy are summarized in Table~\ref{yesno}.

\begin{table}[h]
   \caption{Applicability of the second strategy for the Birkhoff decomposition
            of an XU($n$) matrix with $n=p^w$.}
   \centering
   \vspace{3mm}
\begin{tabular}{|l|ll|}
   \hline
            & $p=2$ & $p \ge 3$ \\
   \hline
   $w=1$    & no    & no        \\
   $w=2$    & yes   & yes       \\
   $w\ge 3$ & no    & yes       \\
   \hline 
\end{tabular}
\label{yesno}
\end{table}

\section{Conclusion}

According to \cite{debaerdemacker}, every unit-linesum $n \times n$ unitary matrix
can be decomposed as a weighted sum of the $n \times n$ permutation matrices,
such that both the sum of the weights and the sum of the squared moduli of the weights
equal unity. Such Birkhoff sum contains $n!$~terms.
In the present paper, we demonstrate the following:
\begin{itemize}
\item If $n \ge 4$, then $n!/2$ terms suffice.
\item If $n=p^w$ with $p$ an arbitrary prime and $w$ an arbitrary integer, then
      $p^w(p^w-p^{w-1})(p^w-p^{w-2})...(p^w-p)(p^w-1)$ suffice.
\item If $n=p^w$ with $p$ an arbitrary odd prime and $w$ an integer $\ge 2$, then
      $p^w(p^w-p^{w-1})(p^w-p^{w-2})...(p^w-p)(p^w-1)/2$ suffice.
\end{itemize}
For numerical examples, see Table~\ref{tabelletje}. 

\begin{table}[h]
   \caption{Number of Birkhoff terms in the decomposition 
            of an arbitrary $n \times n$ unit-linesum unitary matrix.}
   \vspace{3mm}
\begin{tabular}{|c|rrrrrrrrrrr|}
   \hline
   $n$ & 1 & 2 & 3 &  4 &  5 &   6 &  7 &     8 &   9 &        10 &  11 \\
       & 1 & 2 & 6 & 12 & 20 & 360 & 42 & 1,344 & 216 & 1,814,400 & 110 \\
   \hline
\end{tabular}

\begin{tabular}{crrrrrrrrrrr}
       &   &   &   &    &    &     &    &     &     &           &     
\end{tabular}

\begin{tabular}{|c|rrrrrr|}
   \hline
   $n$ &          12 &  13 &             14 &              15 &      16 &  17 \\
       & 239,500,800 & 156 & 43,589,145,600 & 653,837,184,000 & 322,560 & 272 \\
  \hline
\end{tabular}
\label{tabelletje}
\end{table}

The case of $n$ equal to the product of two different primes
is left for further investigation.

\appendix

\section{The group of supercirculant permutation matrices}


The supercirculant $n \times n$ permutation matrices form a group.
Indeed, the product of two such matrices (say $S_{a,x}$ and $S_{b,y}$) yields a third such matrix.
In order to prove this fact, we compute the matrix entry at position $(u,v)$:
\bea
(S_{a,x}\ S_{b,y})_{u,v} & = & \sum_f (S_{a,x})_{u,f}\ (S_{b,y})_{f,v} \\
                      & = & \sum_f \delta_{f,\, a+ux}\ 
                                   \delta_{v,\, b+fy} \\
                      & = & \delta_{v,\, b+(a+ux)y} \\
                      & = & \delta_{v,\, b+ay+uxy} = (S_{b+ay,\, xy})_{u,v} 
\eea
and hence
\begin{equation}
S_{a,x}\ S_{b,y} = S_{b+ay,\, xy} \ .
\label{CCC}
\end{equation}
If $n$ is a prime $p$, each non-zero number~$x$ has an inverse number $x^{-1}$.
Applying (\ref{CCC}), we find
\[
S_{a,x}\ S_{-ax^{-1},\, x^{-1}} = S_{0,1} \ .
\]
The right-hand side being the $p \times p$ unit matrix,
the result proves that each supercirculant matrix has an inverse matrix
that also is supercirculant:
\[
(S_{a,x})^{-1} = S_{-ax^{-1},\, x^{-1}} \ .
\]
We conclude by considering two applications of eqn(\ref{CCC}):
\begin{itemize}
\item
choosing $x=y=1$ leads to
\[
S_{a,1}\ S_{b,1} = S_{a+b,\, 1}
\]
illustrating that the $p$ matrices $S_{a,1}$ are isomorphic to the addition modulo~$p$;
\item
choosing $a=b=0$ leads to
\[
S_{0,x}\ S_{0,y} = S_{0,\, xy}
\]
illustrating that the $p-1$ matrices $S_{0,x}$ are isomorphic to the multiplication modulo~$p$.
\end{itemize}
Each supercirculant matrix can be decomposed
as the product of a zero-shift matrix and a unit-pitch matrix:
\bea
S_{a,x} & = &                  S_{0,x} \, S_{a,1} \\
        & = & S_{ax^{-1},1} \, S_{0,x}            \ .
\eea

\section{The trace of a supercirculant permutation matrix}


We compute the trace of the supercirculant permutation matrix~$S_{a,x}$:
\[
\mbox{Tr}(S_{a,x}) = 
\sum_u   (S_{a,x})_{u,u} = 
\sum_u  \delta_{u,\, a + ux} \ .
\]
If the eqn
\[
u(1-x) = a
\]
is fulfilled, then the corresponding number~$u$ points to 
a unit entry in position $(u,u)$ of the matrix~$S_{a,x}$.
We notice:
\begin{itemize}
\item If $x \neq 1$, then $u = a(1-x)^{-1}$ is the one and only solution; 
\item if $x=1$ and $a \neq 0$, then the eqn has no solution~$u$;
\item if $x=1$ and $a=0$,      then $u$ may have any value from $\{0,\ 1,\ 2,\ ...,\\ p-1\}$.
\end{itemize}
Thus we conclude:
\begin{itemize}
\item Tr$(S_{a,x}) = 1$, if $x \neq 1$,  
\item Tr$(S_{a,1}) = 0$, if $a \neq 0$, and
\item Tr$(S_{0,1}) = p$.
\end{itemize} 

\section{The determinant of a supercirculant permutation matrix}


As mentioned in Appendix~A, each supercirculant matrix
can be decomposed as follows:
\[
S_{a,x} = S_{0,x}\, S_{a,1} \ .
\]
Hence:
\[
\mbox{Det}(S_{a,x}) = \mbox{Det}(S_{0,x})\, \mbox{Det}(S_{a,1}) \ .
\]
We have $S_{a,1}=(S_{1,1})^a$ and therefore $\mbox{Det}(S_{a,1})=(\mbox{Det}(S_{1,1}))^a$.
If $p$ is odd, then $\mbox{Det}(S_{1,1})=1$, such that $\mbox{Det}(S_{a,1})=1$. 
In other words: for odd primes, all of the $p$~circulant permutation matrices have unit determinant.
The situation is different for the $p-1$ matrices~$S_{0,x}$.
Half of them have unit determinant and half of them have determinant
equal to~$-1$. In order to prove this fact,
the key observation is the fact that the cyclic group is Abelian;
so there exists a similarity transformation that diagonalizes {\it all} matrices $S_{0,x}$.
We now prove that the following matrix~$F$ serves our purpose:
\[
F_{u,v} = \left\{ \begin{array}{ll}
                  1 & \mbox{ if } u=v=0 \\

                  0 & \mbox{ if } u=0 \mbox{ and } v \neq 0 \\
                  0 & \mbox{ if } u \neq 0 \mbox{ and } v=0 \\
                  \frac{\omega^{v\fie(u)}}{\sqrt{p-1}}
                    & \mbox{ if } u \neq 0 \mbox{ and } v \neq 0 \ , \end{array} \right. 
\]
where $\omega= \exp(\frac{2\pi i}{p-1})$ is the $(p-1)$~th root of unity,
and the function $\fie (a)$ gives the `position' of the number~$a$ in
the cyclic group ${\bf C}_{p-1}$ (multiplicative group modulo~$p$),
as a power of the (a~priori unknown) generator $g$, i.e.
\[
a = g^{\fie(a)} \ .
\]
From this definition, the following interesting properties of $\fie$
can be deduced:
\bea
\fie(1)  & = & 0 \\
\fie(g)  & = & 1 \\
\fie(ab) & = & \fie(a)+\fie(b) \ . 
\eea
These properties are key in the following derivation.
We compute the similarity transformation given by $F^{\dagger}S_{0,x}F$.
Because both $F$ and $S_{0,x}$ are block diagonal with a single~1
in the upper-left corner, we only need to compute the lower-right part:
\bea
(F^{\dagger}S_{0,x}F)_{u,v} & = & \sum_{k=1}^{p-1} \sum_{l=1}^{p-1} 
                                  \overline{F_{k,u}}\, (S_{0,x})_{k,l}\, F_{l,v}         \\
                            & = & \frac{1}{p-1}\, \sum_{k=1}^{p-1} \sum_{l=1}^{p-1} 
                                  \omega^{-u\fie(k)}\, \delta_{l,xk}\, \omega^{v\fie(l)} \\ 
                            & = & \frac{1}{p-1}\, \sum_{k=1}^{p-1}
                                  \omega^{-u\fie(k)+v\fie(xk)}                           \\
                            & = & \frac{1}{p-1}\, \sum_{k=1}^{p-1}
                                  \omega^{-u\fie(k)+v\fie(x)+v\fie(k)}      \\
                            & = & \omega^{v\fie(x)}\, \delta_{u,v} \ .
\eea
This result leads to two conclusions:
\begin{itemize}
\item By choosing $x=1$, we find that $(F^{\dagger}F)_{u,v} = \delta_{u,v}$ 
      and thus that $F$ is unitary.
\item By choosing $x$ arbitrary, we find that the matrix $S_{0,x}$ has
      the eigenvalues $\omega^{v\fie(x)}$ plus an additional~1
      from the upper-left matrix block. 
\end{itemize}
The determinant is just the product of all eigenvalues:
      \bea
      \mbox{Det}(S_{0,x}) & = & \prod_{v=1}^{p-1} \omega^{v\fie(x)} 
                            =   \omega^{\fie(x)\sum_{v=1}^{p-1} v} = \omega^{\fie(x)\, \frac{p(p-1)}{2}} \\
                          & = & e^{\frac{2\pi i}{p-1}\, \fie(x)\,\frac{p(p-1)}{2}}
                            =   e^{\pi i \fie(x)p} \ .
      \eea
Now, if $p$ is an odd prime, then $e^{\pi ip}=-1$, such that $\mbox{Det}(S_{0,x}) = (-1)^{\fie(x)}$,
which proves that the sign of the determinant of $S_{0,x}$ alternates 
in the chain of successive elements of {\bf C}$_{p-1}$. 
More in particular, the position of $x=g$ always is $\fie(g)=1$, so we have $\mbox{Det}(S_{0,g}) = -1$.

We note that the above results for both $S_{a,1}$ and $S_{0,x}$ are only valid for odd primes~$p$.
If $p$ is even, i.e.\ if $p=2$, 
then there exist only two supercirculant matrices 
$S_{0,1} = \tiny \left( \begin{array}{cc} 1 & 0 \\ 0 & 1 \end{array} \right)$, with determinant equal to~$1$, and 
$S_{1,1} = \tiny \left( \begin{array}{cc} 0 & 1 \\ 1 & 0 \end{array} \right)$, with determinant equal to~$-1$.


\section{The pitch matrix}


In (\ref{ME}),
the epicirculant matrix $E_{{\bf a},{\bf x}}$ needs a unit entry in position 
$({\bf k},{\bf a} + {\bf x}{\bf k})$ if
\[
(M_{r,s})_{{\bf k},{\bf a} + {\bf x}{\bf k}} = (M_{r,s})_{{\bf 0},{\bf a}} 
\]
implying
\[
f(k,r)-f(s,a + \sum_u \sum_v x_{u,v} k_v p^u) = f(0,r)-f(s,a)
\]
or
\[
\sum_j k_jr_j - \sum_j s_j\left(a_j + (\, \sum_u \sum_v x_{u,v} k_v p^u\, )_j\right) = - \sum s_ja_j
\]
and thus
\[
\sum_j s_j \sum_v x_{j,v} k_v = \sum_j k_jr_j
\]
or
\[
\sum_v k_v \sum_j x_{j,v} s_j = \sum_v k_vr_v
\]
and thus
\[
\sum_v k_v\, (\, \sum_j x_{j,v} s_j - r_v \, ) = 0 \ . 
\]
We fulfil this condition by the set of $w$ non-coupled eqns
\begin{equation}
\sum_j s_j x_{j,v}  = r_v \ .
\label{set}
\end{equation}
For each eqn, we expect $p^{w-1}$ solutions
(as we can choose $w-1$ out of the $w$~dits $x_{j,v}$ 
arbitrarily from $\{ 0, 1, ..., p-1\}$).
However, many solutions have to be rejected.
Indeed, each column of the matrix $E_{{\bf a},{\bf x}}$ in (\ref{ME}) 
should contain one and only one unit entry.
For this purpose, it is necessary and sufficient that
the matrix {\bf x} is invertible.
Proof is as follows.
 We require that for any two different row numbers ($k' \neq k$)
 the unit entry of the permutation matrix is in another column:
 \[
 {\bf a} + {\bf xk'} \neq {\bf a} + {\bf xk}  
 \]
and thus ${\bf x}({\bf k'}-{\bf k}) \neq {\bf 0}$.
This requires that for any non-zero number $K$ we have
 \[
 {\bf xK} \neq {\bf 0} \ .
 \]
 This, in turn, requires that the rows of {\bf x} are linearly
 independent and thus that the matrix~{\bf x} is invertible. 

We now prove that, for any pair $(r,s)$,
the set (\ref{set}) has at least one acceptable solution,
i.e.\ a solution such that the matrix {\bf x} is invertible.
Indeed:
\begin{itemize}
\item Because both $r$ and $s$ are non-zero,
      at least one dit $r_u$ is non-zero and
      at least one dit $s_j$ is non-zero.
      Let $r_{\alfa}$ be the least-significant non-zero dit of $r$;
      let $s_{\beta}$ be the least-significant non-zero dit of $s$.
\item We choose all dits $x_{j,v}=0$,
      except the dits $x_{v,v}$, $x_{\beta, v}$, and $x_{\alfa, \beta}$.
      Thus eqns (\ref{set}) become
      \begin{eqnarray}
      s_vx_{v,v}               + s_{\beta}x_{\beta,v}     & = & r_v       \ \mbox{mod}\ p \mbox{ if } v \neq \beta \nonumber \\
      s_{\alfa}x_{\alfa,\beta} + s_{\beta}x_{\beta,\beta} & = & r_{\beta} \ \mbox{mod}\ p               \ . \label{sxr2}
      \end{eqnarray}
\item For $v \neq \alfa$ and $v \neq \beta$, we choose $x_{v,v}=1$.
      Further we choose $x_{\alfa,\alfa}=0$ and $x_{\alfa,\beta}=1$.
      Thus eqns (\ref{sxr2}) become
      \begin{eqnarray}
      s_{\beta}x_{\beta,v}     = r_v       & - \ s_v       & \ \mbox{mod}\ p 
                     \ \mbox{ if }  v \neq \alfa \ \mbox{ and } v \neq \beta \nonumber    \\
      s_{\beta}x_{\beta,\alfa} = r_{\alfa} &               & \ \mbox{mod}\ p \label{sxr3} \\
      s_{\beta}x_{\beta,\beta} = r_{\beta} & - \ s_{\alfa} & \ \mbox{mod}\ p \nonumber
      \end{eqnarray}
      which lead to a single solution set $x_{\beta,v}$. 
\end{itemize}

The resulting pitch matrix~{\bf x} consists of a non-zero diagonal, 
one non-zero row, and one extra unit entry.
E.g.\ for $w=7$, $\alfa=2$, and $\beta=4$, we have:
\[
\left( \begin{array}{ccccccc}
1       &         &         &         &         &         &         \\[1mm]
        & 1       &         &         &         &         &         \\[1mm]
        &         & 0       &         &    1    &         &         \\[1mm]
        &         &         & 1       &         &         &         \\[1mm]
x_{4,0} & x_{4,1} & x_{4,2} & x_{4,3} & x_{4,4} & x_{4,5} & x_{4,6} \\[1mm]
        &         &         &         &         & 1       &         \\[1mm]
        &         &         &         &         &         & 1       \end{array} \right) \ .
\]
We note that here Det({\bf x}) equals $x_{4,2}$.
In general, we have 
\[
\mbox{Det}({\bf x}) = \pm\  x_{\beta,\alfa} = \pm\ r_{\alfa}\, s_{\beta}^{-1}\ .
\]
Because Det({\bf x})~$\neq 0$, we have that {\bf x} is invertible.

\section{The group of epicirculant permutation matrices}


The epicirculant permutation matrices form a group.
An arbitrary entry (at location $({\bf k},{\bf l})$) of such matrix $E_{{\bf a},{\bf x}}$
is $\delta_{{\bf l},\, {\bf a} + {\bf xk}}$.
The product of two such matrices yields a third such matrix. Indeed:
\bea
(E_{{\bf a},{\bf x}}\ E_{{\bf b},{\bf y}})_{u,v} 
& = & \sum_f (E_{{\bf a},{\bf x}})_{u,f}\ (E_{{\bf b},{\bf y}})_{f,v} \\
& = & \sum_f \delta_{{\bf f},\, {\bf a} + {\bf xu}} \   
             \delta_{{\bf v},\, {\bf b} + {\bf yf}} \\
& = & \delta_{{\bf v},\, {\bf b} + {\bf ya} + {\bf yxu}} \\  
& = & (E_{{\bf b} + {\bf ya},\, {\bf yx}})_{u,v}
\eea
and hence
\[
E_{{\bf a},{\bf x}}\ E_{{\bf b},{\bf y}} = E_{{\bf b} + {\bf ya},\, {\bf yx}} \ .
\]
Straightforward application of this result leads to
\[
E_{{\bf a},{\bf x}}\ E_{-{\bf x}^{-1}{\bf a},\, {\bf x}^{-1}} = E_{{\bf 0},{\bf 1}} \ .
\]
The right-hand side being the $p^w \times p^w$ unit matrix,
the result proves that each epicirculant matrix has an inverse matrix
that also is epicirculant:
\[
(E_{{\bf a},{\bf x}})^{-1} = E_{-{\bf x}^{-1}{\bf a},\, {\bf x}^{-1}} \ .
\]

Each epicirculant matrix can be decomposed
as the product of a matrix with zero shift vector {\bf a} and a matrix
with unit pitch matrix {\bf x}: 
\bea
E_{{\bf a},{\bf x}} & = & E_{{\bf 0},{\bf x}}       \, E_{{\bf a},{\bf 1}}  \\
                    & = & E_{{\bf x}^{-1}{\bf a},{\bf 1}} \, E_{{\bf 0},{\bf x}}  \ .
\eea

\section{The trace of an epicirculant permutation matrix}


We compute the trace of the epicirculant permutation matrix~$E_{{\bf a},{\bf x}}$:
\[
\mbox{Tr}(E_{{\bf a},{\bf x}}) = 
\sum_u   (E_{{\bf a},{\bf x}})_{u,u} = 
\sum_u  \delta_{{\bf u},\, {\bf a} + {\bf xu}} \ .
\]
If the eqn
\[
{\bf (1-x)u} = {\bf a}
\]
is fulfilled, then the corresponding number~$u$ points to 
a unit entry in position $(u,u)$ of the matrix~$E_{{\bf a},{\bf x}}$.
Here, {\bf 1} denotes the $w \times w$ unit matrix.
We notice:
\begin{itemize}
\item If ${\bf (1-x)}$ is invertible, 
      then ${\bf u} = {\bf (1-x)}^{-1}{\bf a}$ is the one and only solution;  
\item if ${\bf (1-x)} = {\bf 0}$ and ${\bf a} \neq {\bf 0}$, 
      then the eqn has no solutions~${\bf u}$;
\item if ${\bf (1-x)} = {\bf 0}$ and ${\bf a} = {\bf 0}$, 
      then $u$ may have any value from $\{ 0, 1, 2, ...,\\ p^w-1\}$;
\item if ${\bf (1-x)}$ is neither invertible nor zero, 
      then ${\bf (1-x)}$ has rank $\lambda$ with $1 \le \lambda \le w-1$ and 
      ${\bf u}$ can have as many values as there are
      solutions of the eqn ${\bf (1-x)u} = {\bf 0}$, i.e.\ 
      as the size of the kernel of $({\bf 1}-{\bf x})$, i.e.\ $p^{w-\lambda}$. 
\end{itemize}
Thus we conclude:
\begin{itemize}
\item Tr$(E_{{\bf a},{\bf 1}}) = 0$, if ${\bf a} \neq {\bf 0}$, 
\item Tr$(E_{{\bf 0},{\bf 1}}) = p^w$, and
\item Tr$(E_{{\bf a},{\bf x}}) = p^{w-\lambda}$, if $({\bf 1-x})$ has rank $\lambda \neq 0$.
\end{itemize} 

\end{document}